\documentclass[aps,prl,twocolumn,groupedaddress]{revtex4}

\usepackage{bm}
\usepackage{graphicx}
\usepackage{color}

\begin{document}


\title{Direct Observation of Non-Monotonic $d_{x^2-y^2}$-Wave Superconducting Gap in Electron-Doped High-$T_c$ Superconductor 
Pr$_{0.89}$LaCe$_{0.11}$CuO$_4$}



\author{
		H. Matsui$^1$,
		K. Terashima$^1$,
		T. Sato$^1$,
		T. Takahashi$^1$,
		M. Fujita$^2$, and
		K. Yamada$^2$}
\affiliation{$^1$Department of Physics, Tohoku University, Sendai 980-8578, Japan}
\affiliation{$^2$Institute for Materials Research, Tohoku University, Sendai 980-8577, Japan}


\date{\today}

\begin{abstract}
We performed high-resolution angle-resolved photoemission spectroscopy on electron-doped high-$T_c$ superconductor 
Pr$_{0.89}$LaCe$_{0.11}$CuO$_4$ to study the anisotropy of the superconducting gap.  The observed momentum dependence 
is basically consistent with the $d_{x^2-y^2}$-wave symmetry, but obviously deviates from the monotonic $d_{x^2-y^2}$ gap function.  
The maximum gap is observed not at the zone boundary, but at the hot spot where the antiferromagnetic 
spin fluctuation strongly couples to the electrons on the Fermi surface.  The present experimental results unambiguously indicate 
the spin-mediated pairing mechanism in electron-doped high-$T_c$ superconductors.
\end{abstract}

\pacs{74.72.Jt, 74.25.Jb, 79.60.Bm}

\maketitle
The anisotropy of superconducting (SC) gap is a direct clue for understanding the origin and mechanism of superconductivity.  
It is generally accepted that the SC-gap symmetry of hole-doped high-$T_c$ superconductors (HTSCs) is $d_{x^2-y^2}$ wave and 
is described with the gap function of the monotonic $d_{x^2-y^2}$ form $\Delta$({\it k}) $\propto$ cos($k_x$a)-cos($k_y$a) \cite{Shen,Ding}, 
where the maximum and 
zero SC gaps are located at the Brillouin-zone (BZ) boundary and the diagonal, respectively.  In electron-doped HTSCs, on the 
other hand, the SC-gap symmetry is still under hot debate.  Although a general consensus for the overall $d_{x^2-y^2}$ wave in the 
optimally doped region has been established by the microwave \cite{Koka,Proz}, scanning SQUID microscopy \cite{Tsuei}, and angle-resolved 
photoemission (ARPES) experiments \cite{Sato,Arm}, it has been proposed that the gap function in electron-doped HTSCs is substantially 
deviated from the monotonic $d_{x^2-y^2}$ wave \cite{Blum,Khod,Yoshi} and further may change into a different symmetry such as 
{\it s} wave in the over-doped region \cite{Khod,Skin,Bisw,Pron,Balc}.

These arguments on the SC-gap anisotropy are related to the Fermi surface (FS) geometry with respect to the magnetic BZ.  
If the antiferromagnetic (AF) spin fluctuation mediates the pairing in HTSCs, the SC gap is expected to have a large value at 
particular Fermi momenta ($k_F$) connected to each other by the AF scattering vector {\it Q} = ($\pi$, $\pi$) \cite{Scal}.  This $k_F$ point, 
so-called ``hot spot", is defined as an intersection of the FS and the AF-BZ boundary as shown in Fig. 1.  In the hole-doped case, 
the large circular FS centered at ($\pi$, $\pi$) cuts the AF-BZ boundary very close to ($\pi$, 0), producing the hot spot near ($\pi$, 0).  
This situation does not alter the characteristics of the original monotonic $d_{x^2-y^2}$ gap function with the maximum gap at ($\pi$, 0).  
In contrast, in the electron-doped case the hot spot is moved toward the zone diagonal due to the shrinkage of hole-like FS, which may distort 
the monotonic $d_{x^2-y^2}$ gap function by displacing the maximum gap from ($\pi$, 0) toward ($\pi$/2, $\pi$/2) \cite{Blum,Khod,Yoshi}.  
Furthermore, the proximity of the pairing potential with the opposite sign around the zone diagonal may suppress the $d_{x^2-y^2}$ 
gap symmetry itself \cite{Khod}.  Although the detailed momentum dependence of the SC gap has been well studied by ARPES for 
hole-doped HTSCs \cite{Shen,Ding,Meso,Feng,Bori,Matsui}, that of electron-doped HTSCs has been hardly measured because of the small 
(one-order smaller) SC gap compared with that of hole-doped ones.  However, the experimental elucidation of the gap anisotropy in 
electron-doped HTSCs is highly desired to understand the origin and mechanism of the high-$T_c$ superconductivity.

In this Letter, we report the direct observation of the non-monotonic $d_{x^2-y^2}$ SC gap in the electron-doped HTSC 
Pr$_{0.89}$LaCe$_{0.11}$CuO$_4$ (PLCCO) 
by high-resolution ARPES.  From the detailed measurements of the SC gap along the FS, we found that the maximum SC gap is not 
around the BZ boundary as expected from the monotonic $d_{x^2-y^2}$ gap function, but at the hot spot between ($\pi$, 0) and 
($\pi$/2, $\pi$/2), where the spin fluctuation is expected to most strongly couple to the electrons on the FS.  We discuss the implication of 
the present ARPES results in relation to the origin of the high-$T_c$ superconductivity.

\begin{figure}
\includegraphics[width=3.4in]{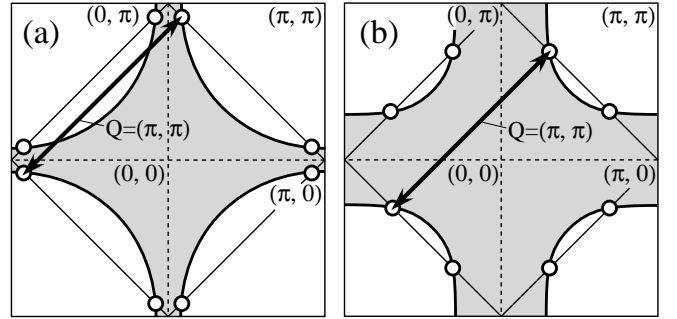}%
\caption{Relation between the Fermi surface and the antiferromagnetic Brillouin zone for (a) hole- and (b) electron-doped HTSCs.  
Thick solid curve and thin straight line show the FS and the AF-BZ, respectively.  Arrow and open circle show the AF scattering 
vector {\it Q} = ($\pi$, $\pi$) and the hot spot,  respectively.}
\end{figure}

High-quality single crystals of PLCCO (optimally doped, transition temperature $T_c$ = 26 K) were grown by the traveling-solvent 
floating-zone method \cite{Fujita}.  ARPES measurements were performed with GAMMADATA-SCIENTA SES-200 spectrometer with a 
high-flux discharging lamp and a toroidal grating monochromator at Tohoku University.  We used the He I$\alpha$ resonance line (21.218 eV) 
to excite photoelectrons.  The energy and angular (momentum) resolutions were set at 5 meV and 0.2$^{\circ}$ (0.01\AA$^{-1}$), respectively.  
A clean surface of sample for ARPES measurements was obtained by {\it in-situ} cleaving along the (001) plane.  The Fermi level of 
sample was referred to that of a gold film evaporated onto the sample substrate.

Figure 2 shows the ARPES spectra near the Fermi level ($E_F$) of PLCCO measured along three representative cuts in the BZ (see the inset) 
and the corresponding intensity plot as a function of the momentum and the binding energy.  We find in Fig. 2 that a highly dispersive 
band crosses $E_F$ in all three cuts, forming a hole-like FS centered at ($\pi$, $\pi$).  However, the spectral line shape looks quite different 
among the three cuts.  In the ($\pi$, 0)-($\pi$, $\pi$) cut (Figs. 2a and 2d), we clearly see the characteristic quasiparticle (QP) behavior near 
$E_F$, namely, the sudden band bending and/or splitting at around 50 meV binding energy, indicative of the QP mass-renormalization 
effect near $E_F$.  In the cut which crosses the hot spot (Figs. 2b and 2e), we find a substantial suppression of the spectral weight 
from $\sim$100 meV to $E_F$, which produces a pseudogap at $E_F$.  It is noted here that this pseudogap is different from the 
so-called small pseudogap, because it has been reported that the small pseudogap is comparable in size to the SC gap \cite{DingPG}.  
In the diagonal cut (Figs. 2c and 2f), the electron correlation effect such as band bending is not seen and an almost straight dispersive band 
crosses $E_F$.  This strong anisotropy of the electronic structure near $E_F$ as a function of momentum observed in PLCCO is consistent 
with the previous ARPES observation on a different electron-doped HTSC Nd$_{2-x}$Ce$_x$CuO$_4$ \cite{ArmPG,MatsuiPG} and is 
well explained in terms of the AF correlation effect.  Since the AF scattering vector Q = ($\pi$, $\pi$) connects the electronic states at 
the AF-BZ boundary, the strong coupling of electrons with the AF fluctuation takes place at the intersection of the band and the 
AF-BZ boundary.  At the hot spot, this intersection is just at $E_F$, producing the (pseudo)gap at $E_F$ as seen in Fig. 2.  The observed 
differences of band dispersions near $E_F$ for other two cuts are well explained in terms of the relative position of the intersection 
with respect to $E_F$ \cite{MatsuiPG}.  Thus, the observed characteristic behavior of the band dispersion near $E_F$ shows the strong 
influence from the AF fluctuation to the electronic structure near $E_F$.

\begin{figure}
\includegraphics[width=3.4in]{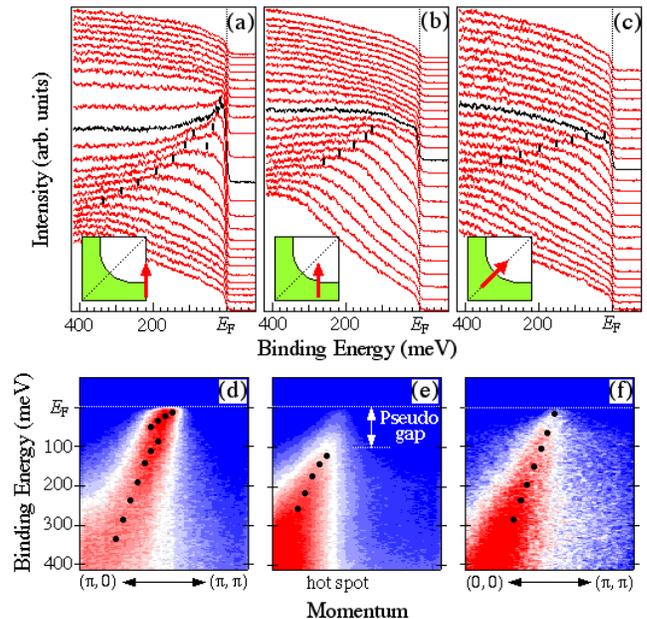}%
\caption{(a)-(c) ARPES spectra of PLCCO along three representative cuts shown by arrow in each inset.  Black-colored spectra 
are measured at the Fermi momentum.  (d)-(f) Corresponding ARPES-intensity plot as a function of the momentum and binding energy, 
showing the experimental band dispersion.  Peak positions in ARPES spectra are shown by bars and dots.}
\end{figure}

Next, we discuss the anisotropy of the SC gap in PLCCO.  Figure 3 shows ARPES spectra in the close vicinity of $E_F$ measured at 
temperatures below and above $T_c$ (8 K and 30 K, respectively) for three different $k_F$ points as shown in the inset.  Points A and C 
are on the ($\pi$, 0)-($\pi$, $\pi$) and the diagonal cut, respectively, and point B corresponds to the hot spot.  We find in Fig. 3 that the 
leading-edge midpoint of the 8-K spectrum is shifted toward the high binding energy with respect to that of the 30-K 
spectrum by a few meV at points A and B, while that of point C does not show such a remarkable temperature-induced 
shift.  This suggests that a $d_{x^2-y^2}$-like SC gap opens at low temperatures in PLCCO.  However, it is remarked that the shift of 
midpoint at point B looks slightly larger than that at point A, exhibiting a striking contrast to the previous ARPES results on 
the hole-doped HTSCs \cite{Shen,Ding,Meso,Feng,Bori,Matsui}.  In order to 
quantitatively estimate the momentum dependence of the SC gap in PLCCO, we numerically fit the ARPES spectra by using the 
phenomenological Fermi-Dirac function with the onset as a free parameter, multiplied by a linear function and convoluted with a 
Gaussian resolution function \cite{Arm}.  Although the shift of leading-edge midpoint ($\Delta$$_{\rm shift}$) in the spectrum is not equal to 
the SC-gap size, it is empirically known that the $\Delta$$_{\rm shift}$ is about a half of the SC gap and serves as a good measure for it 
\cite{Shen,Ding,Sato,Arm,Feng,Bori,SatoLow}.  Estimated 
$\Delta$$_{\rm shift}$'s are 2.0, 2.5, and 0.1 meV with the accuracy of $\pm$0.2 meV at points A, B, and C, respectively.  This clearly indicates 
that the gap function of PLCCO is obviously deviated from the monotonic $d_{x^2-y^2}$ gap function.  We have measured the 
near-$E_F$ ARPES spectrum at 8 K for other several $k_F$ points and estimated the $\Delta$$_{\rm shift}$ value.  Including these points, we plot 
the $\Delta$$_{\rm shift}$'s as a function of the FS angle ($\phi$) in Fig. 4, together with the monotonic $d_{x^2-y^2}$ gap function for comparison.  
The deviation of the measured $\Delta$$_{\rm shift}$ from the monotonic gap function is obvious. The $\Delta$$_{\rm shift}$ is about 2 meV at around 
$\phi$ = 0$^{\circ}$, and gradually {\it increases} on increasing the FS angle, reaching the maximum value of about 2.5 meV at 
$\phi$ = $\sim$ 25$^{\circ}$, which corresponds to the hot spot.  After passing the hot spot, the $\Delta$$_{\rm shift}$ rapidly decreases 
and becomes almost zero at the diagonal ($\phi$ = 45$^{\circ}$).  We have fit this experimental curve $\Delta$$_{\rm shift}$($\phi$) with 
the non-monotonic $d_{x^2-y^2}$ gap function $\Delta$($\phi$) = $\Delta$$_0$[Bcos(2$\phi$)+(1-B)cos(6$\phi$)] which includes the next 
higher harmonic (cos(6$\phi$)) \cite{Blum,Meso,Matsui}.  As shown in Fig. 4, the experimental curve is well fitted with the parameter 
set of $\Delta$$_0$ = 1.9 meV and B = 1.43, indicating the substantial contribution from the second harmonic to the gap function.

\begin{figure}
\includegraphics[width=3.4in]{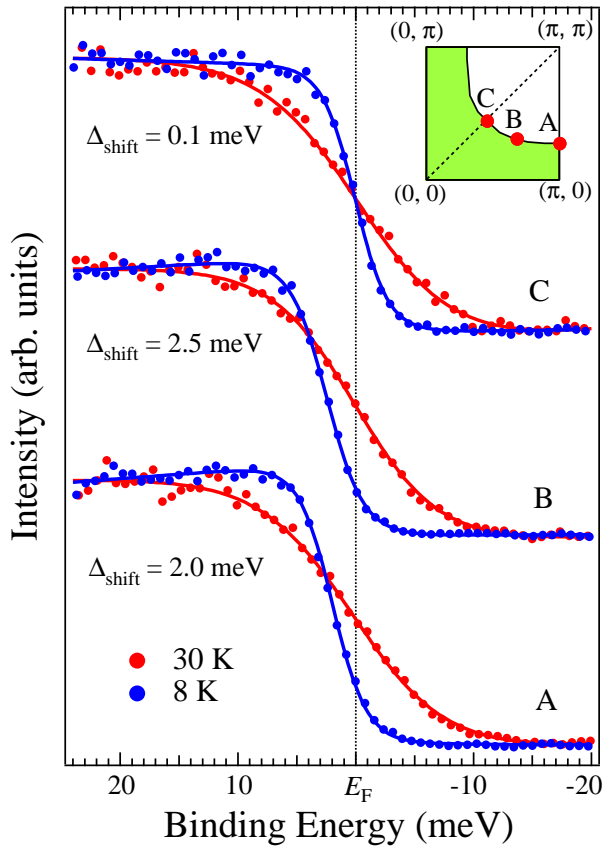}%
\caption{Near-$E_F$ ARPES spectra measured below and above $T_c$ at three $k_F$ points on the FS shown in the inset.  The 8-K and 30-K spectra 
are shown by blue and red dots, respectively.  Solid curves show the fitting of the spectra.}
\end{figure}

\begin{figure}
\includegraphics[width=3.4in]{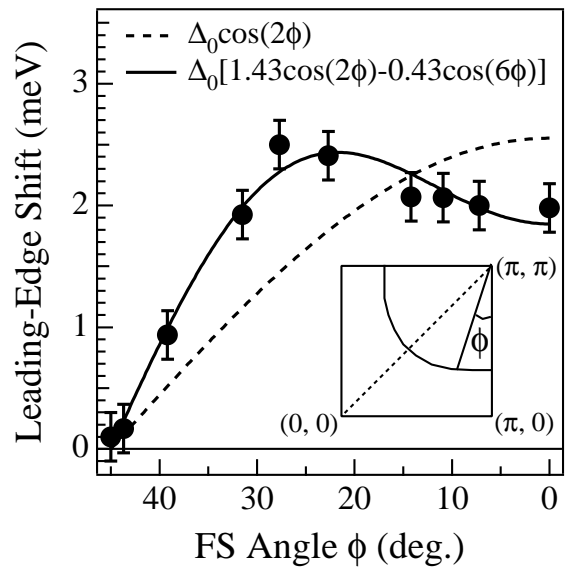}%
\caption{The energy position of leading-edge midpoint ($\Delta$$_{\rm shift}$) plotted as a function of the FS angle ($\phi$).  Solid curve is 
the result of fitting $\Delta$$_{\rm shift}$($\phi$) with the non-monotonic $d_{x^2-y^2}$ gap function, 
$\Delta$$_0$[Bcos(2$\phi$)+(1-B)cos(6$\phi$)], compared with the monotonic $d_{x^2-y^2}$ gap function (broken line).}
\end{figure}

Finally we discuss the present observation in comparison with previous studies.  The polarized Raman spectroscopy \cite{Blum} 
observed that the 2$\Delta$ peak in the $B_{\rm 2g}$ channel (67 cm$^{-1}$) is located at higher frequency than in the $B_{\rm 1g}$ channel 
(50 cm$^{-1}$).  The former and latter Raman channels probe mainly the {\it k} regions around (0, 0)-($\pi$, $\pi$) and ($\pi$, 0), 
respectively. Provided that the 2$\Delta$ 
peak in the $B_{\rm 2g}$ channel is mainly contributed from the hot spot, the Raman experimental result indicates a 1.3-times larger SC 
gap at the hot spot than that around ($\pi$, 0), in good agreement with the present ARPES result (2.5 meV / 2 meV = 1.25).  
It has been theoretically predicted \cite{Khod} that the gap symmetry gradually changes from the $d_{x^2-y^2}$ wave to a different one such as 
{\it s} or {\it p} wave, when the hot spot is moved from ($\pi$, 0) to ($\pi$/2, $\pi$/2) with electron doping.  The calculated gap function in the 
intermediate state exhibits the maximum gap around the hot spot, in good agreement with the present observation in PLCCO.  
The theory has predicted that the transition of gap symmetry occurs when the FS angle of the hot spot ($\phi$$_{hs}$) reaches the critical 
value of $\sim$ 23$^\circ$, which is similar to the $\phi$$_{hs}$ observed in this study, suggesting that the present sample, 
Pr$_{1-x}$LaCe$_x$CuO$_4$ with x = 0.11, is on the boundary of the transition.  ARPES on electron-doped HTSC samples with more 
doping is highly desired to study the transition of the gap symmetry.

In conclusion, the present high-resolution ARPES on the electron-doped HTSC Pr$_{0.89}$LaCe$_{0.11}$CuO$_4$ provides a 
direct evidence for the non-monotonic $d_{x^2-y^2}$-wave SC gap.  The maximum gap is observed not at the BZ boundary as expected 
from the monotonic $d_{x^2-y^2}$ gap function, but at the hot spot between ($\pi$, 0) and ($\pi$/2, $\pi$/2), where the AF spin fluctuation most 
strongly couples to the electrons on the FS.  The experimentally determined gap function is $\Delta$$_0$[1.43cos(2$\phi$)-0.43cos(6$\phi$)], 
indicative of a substantial contribution from the second harmonic of the $d_{x^2-y^2}$ order parameter.  The present results indicate that 
the pairing potential in both electron- and hole-doped HTSCs commonly have the maximal magnitude at the hot spot, suggesting the 
universality of the spin-mediated pairing mechanism in both types of HTSCs.

This work was supported by a grant from the MEXT of Japan.  H.M. thanks a financial support from JSPS.


\newpage


\end{document}